\documentstyle[prl,twocolumn,aps,psfig]{revtex}

\begin{document}
\draft
\preprint{}
\title{On Keller Theorem for Anisotropic Media}
\author{Leonid G. Fel}
\address{School of
Physics and Astronomy, Raymond and Beverly Sackler Faculty of Exact
Sciences\\Tel Aviv University,
Tel Aviv 69978, Israel
\\ e-mail: lfel@post.tau.ac.il
}
\date{\today}
\maketitle

\def\be{\begin{equation}}
\def\ee{\end{equation}}
\def\p{\prime}
\subsection*{Abstract}
The Keller theorem in the problem of effective conductivity in anisotropic 
two-dimensional (2D) many-component composites makes it possible to 
establish a simple inequality $\sigma_{{\sf is}}^e(\sigma^{-1}_i)\cdot
\sigma_{{\sf is}}^e(\sigma_k)> 1$ for the isotropic part 
$\sigma_{{\sf is}}^e(\sigma_k)$ of the 2-nd rank symmetric tensor 
${\widehat \sigma}_{i,j}^e$ of effective conductivity.

\noindent

\widetext
\narrowtext

The extension of the Keller theorem \cite{keller64} in the problem of effective 
conductivity in the infinite 2D two-component composites on the 
many-component case \cite{felberg20} poses a new question on the duality relation for 
the 2-nd rank symmetric tensor ${\widehat \sigma}_{i,j}^e$ of effective conductivity 
in anisotropic media. It is related to the restrictions imposed on the 
linear invariant of ${\widehat \sigma}_{i,j}^e$ which is called an isotropic 
part $\sigma_{\sf is}^e(\sigma_k)$ of effective conductivity. Recently the 
perturbation theory for the infinite 
periodic three-component 2D checkerboard with two-fold 
rotation lattice symmetry was developed \cite{Khalat20} where the coincidence 
of $\sigma_{\sf is}^e(\sigma_k)$ with solution $\sigma_B(\sigma_k)$ of 
Bruggemann 
Eqn was established up to the 6-th order term. This fact is curious because it 
gives grounds to think that Effective Medium Approximation (EMA) describes exactly
$\sigma_{\sf is}^e(\sigma_k)$ in this certain structure.
Here we will discuss this conclusion.

Let us define the isotropic part of conductivity tensor
\begin{equation}
\sigma_{\sf is}^e(\sigma_k)=\frac{1}{2}\;{\sf Tr}\;\;
{\widehat \sigma}_{i,j}^e(\sigma_k)\;,\;\;k=1,2,...,n\;,
\label{keller1}
\end{equation}
which is an invariant scalar with respect to the plane rotation and recall
the Keller theorem for the principal values ${\widehat \sigma}_e^{{\sf xx}},\;
{\widehat \sigma}_e^{{\sf yy}}$ of diagonalized matrix 
${\widehat \sigma}_e^{ij}$ for 2D $n$-component composite
\begin{eqnarray}
&&{\widehat \sigma}_e^{{\sf xx}}(\sigma^{-1}_1,\sigma^{-1}_2,...,\sigma^{-1}_n)\cdot
{\widehat \sigma}_e^{{\sf yy}}(\sigma_1,\sigma_2,...,\sigma_n)=1\;,\nonumber \\
&&{\widehat \sigma}_e^{{\sf yy}}(\sigma^{-1}_1,\sigma^{-1}_2,...,\sigma^{-1}_n)\cdot
{\widehat \sigma}_e^{{\sf xx}}(\sigma_1,\sigma_2,...,\sigma_n)=1\;.\label{keller2}
\end{eqnarray}
Both (\ref{keller1}) and (\ref{keller2}) make us possible to derive 
a simple inequality for $\Lambda_{{\sf is}}^e=\sigma_{{\sf is}}^e
(\sigma^{-1}_i)\cdot\sigma_{{\sf is}}^e(\sigma_k)$
\begin{eqnarray}
\Lambda_{{\sf is}}^e&=&
\frac{1}{4}\left[2+{\widehat \sigma}_e^{{\sf xx}}(\sigma_k)\cdot 
{\widehat \sigma}_e^{{\sf xx}}(\sigma_k^{-1})+
{\widehat \sigma}_e^{yy}(\sigma_k)\cdot{\widehat \sigma}_e^{{\sf 
yy}}(\sigma_k^{-1})\right]=
\nonumber \\ 
&&\frac{1}{4} \left[2+\frac{{\widehat \sigma}_e^{{\sf xx}}(\sigma_k)}
{{\widehat \sigma}_e^{{\sf yy}}(\sigma_k)}+\frac{{\widehat 
\sigma}_e^{{\sf yy}}(\sigma_k)}{{\widehat \sigma}_e^{{\sf 
xx}}(\sigma_k)}\right]\geq 1\;,
\label{keller3}
\end{eqnarray}
where the only isotropic media ${\widehat \sigma}_e^{{\sf xx}}={\widehat \sigma}_e^{{\sf yy}}$
 corresponds to the equality in (\ref{keller3}). At the same time another
isotropic invariant $\Delta_{{\sf is}}^e=
\det {\widehat \sigma}_{ij}^e(\sigma_k)$ satisfies the duality relation
$$
\Delta_{{\sf is}}^e(\sigma_k)\cdot \Delta_{{\sf is}}^e(\sigma_k^{-1})=1\;.
$$
The EMA theory of the infinite 2D $n$-component isotropic comosite has its 
consequence the Bruggemann Eqn \cite{brug35}
\begin{equation}
\sum_{k=1}^n\;\frac{\sigma_k-\sigma_B(\sigma_k)}
{\sigma_k+ \sigma_B(\sigma_k)}=0\;,
\label{brug1}
\end{equation}
which necessarely leads to the duality relation 
\begin{equation}
\sigma_B(\sigma_k^{-1})\cdot \sigma_B(\sigma_k)=1
\label{brug2}
\end{equation}
that reflects both the conformal invariance of the Maxwell Eqns in 2D 
isotropic comosite and $S_n$-permutation invariance of the $n$-colour tessellation 
of the plane. The latter means that $\sigma_{\sf is}^e$ can satisfy the Bruggemann 
Eqn only for isotropic $S_n$-permutation invariant media: 
${\widehat \sigma}_e^{xx}={\widehat \sigma}_e^{yy},\; 
{\widehat \sigma}_e^{xy}=0$ in any reference frame $\{x,y\}$.

The infinite periodic 2D three-component checkerboard was considered in 
\cite{Khalat20} for  symmetrically related partial conductivities 
$(\sigma_1=1,\sigma_{2,3}=1\pm \delta)$. 
\begin{figure}[h]
\centerline{\psfig{figure=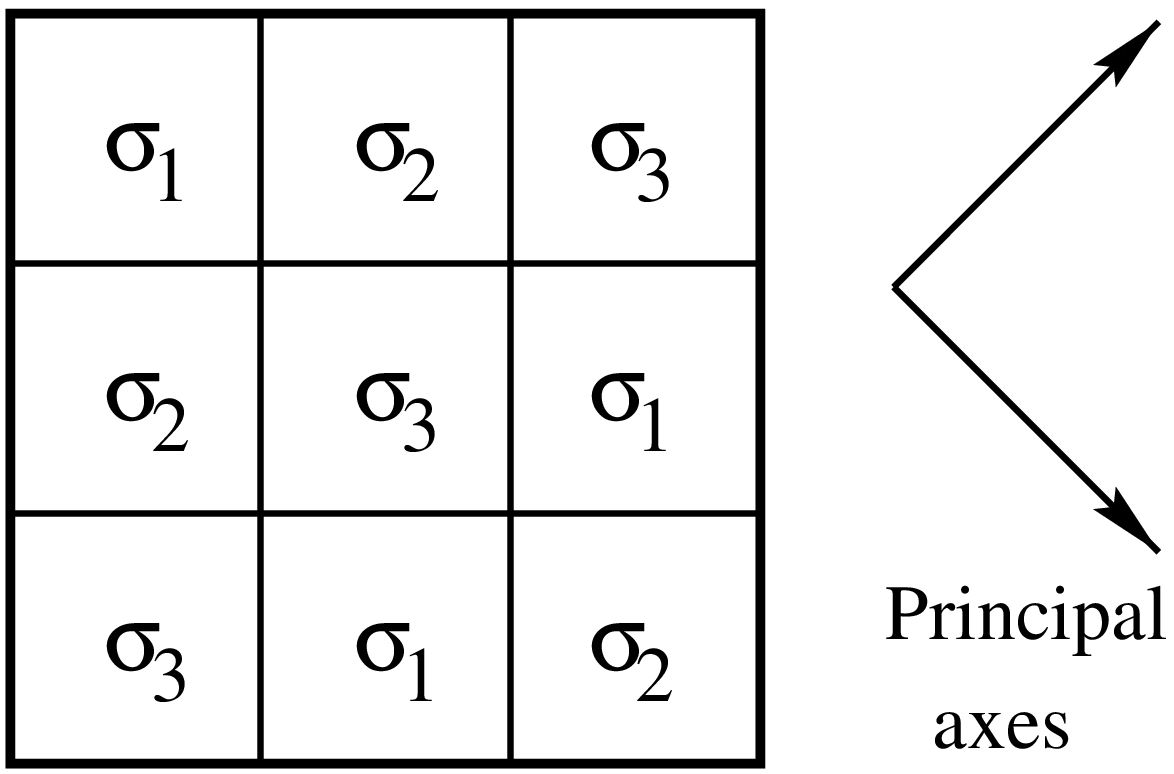,height=3cm,width=5cm}}
\label{kelf1}
\end{figure}
Such structure doesn't possess an isotropy of the 2-nd rank conductivity tensor 
${\widehat \sigma}_e^{i,j}$ that follows from the simple crystallographycal 
consideration \cite{herm34} as well as from the straightforward calculation 
\cite{Khalat20} of the non-diagonal term ${\widehat \sigma}_e^{xy}\propto 
\delta^2$. Therefore $\sigma_{\sf is}^e(\sigma_k)$ for this structure can not 
satisfy the Bruggemann Eqn (\ref{brug1}) even if its coincidence with 
$\sigma_B(\sigma_k)$ riched the $\delta^6$ term in the perturbation theory.


\end{document}